# A NEW REFINED THEORY OF PLATES WITH TRANSVERSE SHEAR DEFORMATION FOR MODERATELY THICK AND THICK PLATES

J.M. MARTÍNEZ VALLE[†]

† *Mechanics Department, EPS; Leonardo da Vinci Building, Rabanales Campus, Cordoba University, 14071, Cordoba, Spain.*
*jmvalle@uco.es*

*Abstract*— **In this paper we propose a new refined shear deformation plate theory which possesses a series of desirable features, the most salient of which are as follows: (i) The loads, which are generally considered to be applied on the middle surface of the plate, act on the upper surface of the plate; (ii) The equations are applicable to the calculation of the stresses in isotropic plates and provide the same order of accuracy as several theories with second order shear deformation effects; (iii) It constitutes a theory, in the sense defined by Love, since it gives easy expressions for application to problems in different fields in architecture and civil engineering.**

*Keywords*— **thick plates, first order shear deformation theory, moderately thick plates.**

## I. INTRODUCTION

A rectangular plate is usually considered thin if its thickness is smaller than a tenth of the minor dimension. When this limitation is not fulfilled one is entering the field of moderately thick plates (a term introduced by Love (1944)), or thick plates.

The current trend in the study of plates can be deduced from the themes of the articles collected by Voyiadjis and Karamanlidis (1990) and Kienzler *et al.* (1990). In the first publication it can be seen that four papers make a direct reference to moderately thick plates. The second one discusses common roots of different new plate and shell theories, review the current state of the art and higher-order shear deformation theories.

Since Reissner (1945), Mindlin (1951), Hencky (1947), and Reismann (1988) elaborated their technical calculation theories (first order shear deformation theory), with the objective of widening the field of application of plates, the theory of bending of moderately thick plates and thick plates (second order plate theories) has been studied by Donnell (1976), Kromm (1953), Panc (1975), Muhammad *et al.* (1990), Voyiadjis and Karamanlidis (1990), and Kienzler (2004), among others.

Generally, all of those theories are characterized by the high level of mathematical complexity required to obtain solutions. Furthermore, the problems which are solved analytically only constitute several specific examples.

In this paper, we present a new shear deformation theory whose main features are presented in the next paragraphs.

## II. HYPOTHESIS AND OBTAINING OF THE GOVERNING EQUATIONS

The hypotheses to consider are as follows:
1) The loads, which are distributed, act on the upper surface of the plate and will be perpendicular to the middle surface, and the displacements of the points located in the middle surface are also sensibly perpendicular to the mentioned middle surface (the middle surface being practically inelastic), that is, $\hat{u} \approx 0$ and $\hat{v} \approx 0$ where $\hat{u}$ and $\hat{v}$ are the displacements according to the x- and y-axes of the points located in the middle surface. The deflection w according to the z-axis of a generic point not located in the middle surface is given by

$$w = \hat{w} + f(x), \quad (1)$$

where $\hat{w}$ is the deflection according to the z-axis of the points located in the middle surface. In the longitudinal deformation calculation subject to the thickness, Poisson's effect is not considered; that is,

$$\varepsilon_x \approx \frac{\sigma_x}{E} \quad (2)$$

2) The *fibres pertaining to the plate, which are straight and perpendicular to the middle surface before the deformation*, do not continue to be perpendicular, and they bend in such a way that the shearing strains are defined by a parabolic distribution throughout the thickness, given by

$$\gamma_{xz} = \left(1 - \frac{4z^2}{h^2}\right)\hat{\gamma}_{xz}; \gamma_{yz} = \left(1 - \frac{4z^2}{h^2}\right)\hat{\gamma}_{yz} \quad (3)$$

where $\hat{\gamma}_{xz}$ and $\hat{\gamma}_{yz}$ are the shearing strains in the points of the middle surface.
3) The rotation $w_{xy}$ of a differential element around the z-axis is null for all points of the plate,

$$w_{xy} = \frac{1}{2}\left(\frac{\partial v}{\partial y} - \frac{\partial u}{\partial x}\right) = 0. \quad (4)$$

This important condition is deduced analytically if we establish the equilibrium of a plate element in its deformed configuration and taking into consideration Reissner´s kinematic assumptions, Martínez Valle (2012).

One denominates by $\hat{w}$ the deflection, according to the z-axis, of *the generic point m located on the plate's mid surface*, by $\vartheta_x$ the angle rotated by the rectilinear segment normal to middle surface around the ox-axis, and by $\vartheta_y$ the angle rotated around the oy-axis. According to its definition, the *shearing strain* in the xz-surface *at the point m located on the middle surface* ($\hat{\gamma}_{xz}$) is

$$\hat{\gamma}_{xz} = \frac{\pi}{2} - \alpha_f, \qquad (5)$$

which, as we can see in Fig. 1, is

$$\hat{\gamma}_{xz} = \vartheta_y + \frac{\partial \hat{w}}{\partial x}. \qquad (6)$$

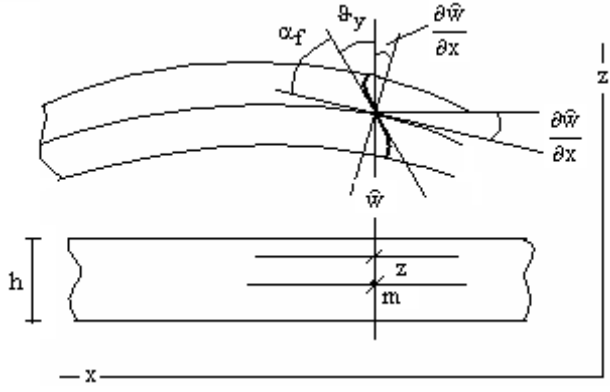

Figure 1: Bending deformation of the plate. Reissner assumption.

Likewise, we deduce

$$\hat{\gamma}_{yz} = -\vartheta_x + \frac{\partial \hat{w}}{\partial y}, \qquad (7)$$

and we may also write

$$\hat{\gamma}_{xz} = \frac{\hat{\tau}_{xz}}{G} \text{ and } \hat{\gamma}_{yz} = \frac{\hat{\tau}_{yz}}{G}, \qquad (8)$$

where $\hat{\tau}_{xz}$ and $\hat{\tau}_{yz}$ are the shearing stresses at *the points located on the middle surface*.

Therefore shearing stresses at the points of the middle surface are

$$\frac{\hat{\tau}_{xz}}{G} = \vartheta_y + \frac{\partial \hat{w}}{\partial x} \; ; \; \frac{\hat{\tau}_{yz}}{G} = -\vartheta_x + \frac{\partial \hat{w}}{\partial y}, \qquad (9)$$

The shearing strains at a generic point are

$$\hat{\gamma}_{xz} = \frac{\partial w}{\partial x} + \frac{\partial u}{\partial z} = \left(1 - \frac{4z^2}{h^2}\right) \hat{\gamma}_{xz}, \qquad (10)$$

$$\hat{\gamma}_{yz} = \frac{\partial w}{\partial y} + \frac{\partial v}{\partial z} = \left(1 - \frac{4z^2}{h^2}\right) \hat{\gamma}_{yz} = -\vartheta_x + \frac{\partial \hat{w}}{\partial y} - \frac{4z^2}{h^2}\frac{\hat{\tau}_{yz}}{G}, \qquad (11)$$

where

$$w'_x = \hat{w}'_x \; , \; w'_y = \hat{w}'_y . \qquad (12)$$

One may deduce the displacements of a generic point:

$$u = \hat{u} + \vartheta_y \cdot z - \frac{4z^3}{3h^2}\frac{\hat{\tau}_{xz}}{G} \; ; v = \hat{v} + \vartheta_x \cdot z - \frac{4z^3}{3h^2}\frac{\hat{\tau}_{yx}}{G} \qquad (13)$$

and from them we obtain the rotations around the x- and y-axes

$$w_{xz} = -\frac{1}{2}(\vartheta_y - \frac{4z^2}{h^2}\frac{\hat{\tau}_{xz}}{G} - \frac{\partial \hat{w}}{\partial x});$$
$$w_{yx} = -\frac{1}{2}(-\vartheta_x - \frac{4z^2}{h^2}\frac{\hat{\tau}_{yz}}{G} - \frac{\partial \hat{w}}{\partial y}). \qquad (14)$$

Considering the third hypothesis and the value of the rotation around the z-axis at all points,

$$w_{xy} = 0 = \hat{w}_{xy} - \frac{z}{2}\left(\frac{\partial \vartheta_y}{\partial y} + \frac{\partial \vartheta_x}{\partial x}\right) + \frac{2z^3}{3h^2 G}\left(\frac{\hat{\tau}_{xz}}{\partial y} - \frac{\hat{\tau}_{yx}}{\partial x}\right) = \hat{w}_{xy} - \frac{z}{2}\left(1 - \frac{4z^2}{3h^2}\right)\left(\frac{\partial \vartheta_y}{\partial y} + \frac{\partial \vartheta_x}{\partial x}\right), \qquad (15)$$

we deduce

$$\frac{\partial \vartheta_y}{\partial y} + \frac{\partial \vartheta_x}{\partial x} = 0. \qquad (16)$$

The strains $\varepsilon_x$, $\varepsilon_{xy}$, and $\gamma_{xy}$ are

$$\varepsilon_x = \hat{\varepsilon}_x + \frac{\partial \vartheta_y}{\partial x} z - \frac{4z^3}{3Gh^2}\frac{\hat{\tau}_{xz}}{\partial x};$$

$$\varepsilon_y = \hat{\varepsilon}_{xy} + \frac{\partial \vartheta_x}{\partial y} z - \frac{4z^3}{3Gh^2}\frac{\hat{\tau}_{yz}}{\partial y} \qquad (17)$$

$$\gamma_{xy} = \hat{\gamma}_{xy} - \frac{\partial \vartheta_x}{\partial x} z - \frac{4z^3}{3Gh^2}\frac{\widehat{\partial \tau}_{yz}}{\partial x} + \frac{\partial \vartheta_y}{\partial y} z - \frac{4z^3}{3Gh^2}\frac{\widehat{\partial \tau}_{xz}}{\partial y} = \hat{\gamma}_{xy} + \left(z - \frac{4z^3}{3h^2}\right)\left(\frac{\partial \vartheta_y}{\partial y} - \frac{\partial \vartheta_x}{\partial x}\right) - \frac{8z^3}{3h^2}\frac{\partial^2 w}{\partial x \partial y} \qquad (18)$$

The shearing stresses $\tau_{xz}$ and $\tau_{yz}$ are deduced from the expressions given by Hooke's law and Lamé's equations:

$$\tau_{xz} = \left(1 - \frac{4z^2}{h^2}\right) \hat{\tau}_{xz}, \qquad (19)$$

$$\tau_{yz} = \left(1 - \frac{4z^2}{h^2}\right) \hat{\tau}_{yz}. \qquad (20)$$

The equilibrium equations of the plate element of differential sides are

$$\frac{\partial Q_{xz}}{\partial x} + \frac{\partial Q_{yz}}{\partial y} + P = 0, \qquad (21)$$

$$Q_{xz} = \frac{\partial M_x}{\partial x} + \frac{\partial M_{xy}}{\partial y}, \qquad (22)$$

$$Q_{yz} = \frac{\partial M_{xy}}{\partial x} + \frac{\partial M_y}{\partial y}. \qquad (23)$$

The transverse shear forces in the faces of the plate element are

$$Q_{xz} = \int_{-\frac{h}{2}}^{\frac{h}{2}} \tau_{xz} \, dz = \int_{-\frac{h}{2}}^{\frac{h}{2}} \left(1 - \frac{4z^2}{h^2}\right) \hat{\tau}_{xz} \, dz = \frac{2h}{3}\hat{\tau}_{xz}, \qquad (24)$$

$$Q_{yz} = \int_{-\frac{h}{2}}^{\frac{h}{2}} \tau_{yz} \, dz = \int_{-\frac{h}{2}}^{\frac{h}{2}} \left(1 - \frac{4z^2}{h^2}\right) \hat{\tau}_{yz} \, dz = \frac{2h}{3}\hat{\tau}_{yz} \qquad (25)$$

and substituting in Eq.21 we obtain

$$\frac{\partial \hat{\tau}_{xz}}{\partial x} + \frac{\partial \hat{\tau}_{yz}}{\partial y} = \frac{-3P}{2h}, \qquad (26)$$

And also after substituting $\hat{\tau}_{xz}$ and $\hat{\tau}_{yz}$ with their values,

$$\frac{\partial \vartheta_y}{\partial x} - \frac{\partial \vartheta_x}{\partial y} + \Delta w = -\frac{-3}{2Gh}P. \qquad (27)$$

The normal stress $\sigma_z$ is determined by means of the following equation of the internal equilibrium of the elasticity:

$$\frac{\partial \tau_{xz}}{\partial x} - \frac{\partial \tau_{yz}}{\partial y} + \frac{\partial \sigma_z}{\partial z} = 0, \quad (28)$$

from which we deduce

$$\frac{\partial \sigma_z}{\partial z} = -\frac{\partial \tau_{xz}}{\partial x} - \frac{\partial \tau_{yz}}{\partial y} = -\left(1 - \frac{4z^2}{h^2}\right)\left(\frac{\partial \hat{\tau}_{xz}}{\partial x} + \frac{\partial \hat{\tau}_{yz}}{\partial y}\right) =$$
$$= \frac{3P}{2h}\left(1 - \frac{4z^2}{h^2}\right) \quad (29)$$

Integrating throughout the thickness between a generic point located at a height z and the upper surface, in which $\sigma_z$ is equal to P, we get

$$P - \sigma_z = \frac{3P}{2h}\int_z^{\frac{h}{2}}\left(1 - \frac{4z^2}{h^2}\right)dz = \frac{3P}{2h}\left(\frac{h}{3} - z + \frac{4z^3}{3h^2}\right). \quad (30)$$

Thus we deduce

$$\sigma_z = \frac{P}{2} + \frac{3Pz}{2h}\left(1 - \frac{4z^2}{3h^2}\right). \quad (31)$$

This last equation is exactly the same which appears in Kromm´s refined plate theory, Panc (1975).
Now we must verify that in the lower surface, where applied loads do not exist, the following is fulfilled:

$$(\sigma_z)_{z=-\frac{h}{2}} = 0, \quad (32)$$

and we find that it is indeed fulfilled. The normal stresses $\sigma_x$ and $\sigma_y$ are deduced from the expressions given by Hooke's law and Lamé's equations:

$$\sigma_z = \frac{E}{1-\mu^2}(\varepsilon_x + \mu\varepsilon_y) + \frac{\mu\sigma_x}{1-\mu} = \frac{E}{1-\mu^2}\left(\hat{\varepsilon}_x + \frac{\partial \vartheta_y}{\partial x}z - \frac{4z^3}{3Gh^2}\frac{\partial \hat{\tau}_{xz}}{\partial x} + \mu\left(\hat{\varepsilon}_y + \frac{\partial \vartheta_x}{\partial y}z - \frac{4z^3}{3Gh^2}\frac{\partial \hat{\tau}_{yz}}{\partial y}\right)\right) + \frac{\mu\sigma_x}{1-\mu} =$$
$$\frac{E}{1-\mu^2}\left(\hat{\varepsilon}_x + \frac{\partial \vartheta_y}{\partial x}z - \frac{4z^3}{3Gh^2}\frac{\partial \hat{\tau}_{xz}}{\partial x} + \mu\left(\hat{\varepsilon}_y + \frac{\partial \vartheta_x}{\partial y}z - \frac{4z^3}{3Gh^2}\frac{\partial \hat{\tau}_{yz}}{\partial y}\right)\right) + \frac{\mu}{1-\mu}\left(\frac{P}{2} + \frac{3Pz}{2h}\left(1 - \frac{4z^2}{3h^2}\right)\right) \quad (33)$$

Now we impose that the in-plane normal force $N_x$ must be null:

$$N_x = \int_{-\frac{h}{2}}^{\frac{h}{2}}\sigma_{xz}\,dz = 0, \quad (34)$$

and it yields

$$\hat{\varepsilon}_x + \mu\hat{\varepsilon}_y = -\frac{\mu(1+\mu)P}{2E} \quad (35)$$

and therefore the normal stress $\sigma_x$ is

$$\sigma_x = \frac{Ez}{1-\mu^2}\left(\frac{\partial \vartheta_y}{\partial x} - \mu\frac{\partial \vartheta_x}{\partial y}\right) - \frac{4z^3}{3Gh^2}\frac{E}{1-\mu^2}\left(\frac{\partial \hat{\tau}_{xz}}{\partial x} + \mu\frac{\partial \hat{\tau}_{yz}}{\partial y}\right) +$$
$$+ \frac{3\mu Pz}{2(1-\mu)h}\left(1 - \frac{4z^2}{3h^2}\right) \quad (36)$$

Similarly we deduce for the normal stress $\sigma_y$

$$\sigma_y = \frac{Ez}{1-\mu^2}\left(\frac{\partial \vartheta_x}{\partial y} - \mu\frac{\partial \vartheta_y}{\partial x}\right) - \frac{4z^3}{3Gh^2}\frac{E}{1-\mu^2}\left(\frac{\partial \hat{\tau}_{yz}}{\partial y} + \mu\frac{\partial \hat{\tau}_{xz}}{\partial x}\right) +$$
$$+ \frac{3\mu Pz}{2(1-\mu)h}\left(1 - \frac{4z^2}{3h^2}\right) \quad (37)$$

with

$$\hat{\varepsilon}_y + \hat{\varepsilon}_x = -\frac{\mu(1+\mu)P}{2E} \quad (38)$$

Eqs.35 and 38 allow us to verify the coherence of the presented theory, observing the fulfilment of the starting hypotheses, and to verify that the normal strains in the middle surface, $\hat{\varepsilon}_x$ and $\hat{\varepsilon}_y$, are very small and lead to $\hat{u} \approx 0, \hat{v} \approx 0$ and

$$\sigma_z = E\varepsilon_z - \frac{P\mu^2}{(1-\mu)} + \frac{2\mu^2}{(1-\mu)}\sigma_z \quad (39)$$

in which he third addend is small compared to $\sigma_z$

$$\sigma_z\left(1 - \frac{2\mu^2}{(1-\mu)}\sigma_z\right) = E\varepsilon_z - \frac{P\mu^2}{(1-\mu)} \quad (40)$$

and the second one,

$$\frac{P\mu^2}{(1-\mu)}, \quad (41)$$

according to Eq.31, is small compared to the other addend of $\sigma_z$, and thus we get $\sigma_z \approx E\varepsilon_z$.
The shearing stress $\tau_{xy}$ is deduced from

$$\tau_{xy} = G\gamma_{xy} = G\left(\frac{\partial u}{\partial y} + \frac{\partial v}{\partial x}\right). \quad (42)$$

Substituting and considering the notes presented in the previous paragraph, we obtain

$$\tau_{xy} = \hat{\tau}_{xy} + Gz\left(\frac{\partial \vartheta_y}{\partial y} + \frac{\partial \vartheta_x}{\partial x}\right) - \frac{4z^3}{3Gh^2}\left(\frac{\hat{\tau}_{xz}}{\partial y} + \frac{\hat{\tau}_{yz}}{\partial x}\right). \quad (43)$$

After integrating throughout the thickness, for example,

$$M_x = \int_{-\frac{h}{2}}^{\frac{h}{2}}\sigma_x z\,dz, \quad (44)$$

the moment stress resultants are expressed by

$$M_x = D\left(\frac{\partial \vartheta_y}{\partial y} + \mu\frac{\partial \vartheta_x}{\partial x}\right) - \frac{D}{5G}\left(\frac{\partial \hat{\tau}_{xz}}{\partial x} + \mu\frac{\widehat{\partial \tau_{yz}}}{\partial y}\right) + \frac{\mu Ph^2}{10(1-\mu)}, (45)$$

$$M_y = D\left(-\frac{\partial \vartheta_x}{\partial y} + \mu\frac{\partial \vartheta_y}{\partial x}\right) - \frac{D}{5G}\left(\frac{\partial \hat{\tau}_{yz}}{\partial y} + \mu\frac{\widehat{\partial \tau_{xz}}}{\partial x}\right) + \frac{\mu Ph^2}{10(1-\mu)}, (46)$$

$$M_{xy} = -\frac{1-\mu}{2}D\left(\frac{\partial \vartheta_y}{\partial y} + \frac{\partial \vartheta_x}{\partial x}\right) - \frac{(1-\mu)D}{10G}\left(\frac{\widehat{\partial \tau_{xz}}}{\partial y} + \frac{\widehat{\partial \tau_{yz}}}{\partial x}\right). \quad (47)$$

The differential equations to determine $\vartheta_x, \vartheta_y$ and $\widehat{w}$ may be obtained by applying the principle of minimum energy and making null the first variation of the total potential energy or alternatively by deciding to apply variational formulation or to raise the static equilibrium equations of the plate element of differential sides. Taking the last method, we substitute the moments calculated before in the last two equations

$$\Delta\vartheta_y + \frac{1+\mu}{2}\frac{\partial}{\partial y}\left(\frac{\partial \vartheta_x}{\partial x} + \frac{\partial \vartheta_y}{\partial y}\right) - \frac{1}{4}\frac{\partial}{\partial x}(\Delta\widehat{w}) = \frac{5(1-\mu)}{h^2}\left(\vartheta_y + \frac{\partial \widehat{w}}{\partial x}\right) -$$
$$- \frac{\mu h^2}{8D(1-\mu)}\frac{\partial P}{\partial x}, \quad (48)$$

$$\Delta\vartheta_x + \frac{1+\mu}{2}\frac{\partial}{\partial x}\left(\frac{\partial \vartheta_y}{\partial y} + \frac{\partial \vartheta_x}{\partial x}\right) - \frac{1}{4}\frac{\partial}{\partial y}(\Delta\widehat{w}) =$$
$$= \frac{5(1-\mu)}{h^2}\left(\vartheta_x + \frac{\partial \widehat{w}}{\partial y}\right) + \frac{\mu h^2}{8D(1-\mu)}\frac{\partial P}{\partial y}, \quad (49)$$

which, according to Eq.16, are reduced to

$$\Delta\vartheta_y - \frac{1}{4}\frac{\partial}{\partial x}(\Delta\widehat{w}) = \frac{5(1-\mu)}{h^2}\left(\vartheta_y + \frac{\partial \widehat{w}}{\partial x}\right) - \frac{\mu h^2}{8D(1-\mu)}\frac{\partial P}{\partial x},$$
$$(50)$$

$$\Delta\vartheta_x + \frac{1}{4}\frac{\partial}{\partial y}(\Delta\widehat{w}) = \frac{5(1-\mu)}{h^2}\left(\vartheta_x - \frac{\partial\widehat{w}}{\partial y}\right) + \frac{\mu h^2}{8D(1-\mu)}\frac{\partial P}{\partial y}. \tag{51}$$

These equations, along with Eq.27 constitute a **calculation system of connected differential equations**, in the sense of simultaneously carrying out the calculation of displacements and rotations that allow us to determine $\vartheta_x$, $\vartheta_y$ and $\widehat{w}$ if we consider the boundary conditions of the plate.

### III. GOVERNING EQUATIONS DISCONNECTED FOR DISPLACEMENTS AND ROTATIONS

The system formed by Eqs.27, 51, and 52 may be disconnected, that is, to separate the calculation of displacements from the calculation of rotations, if we transform it applying Laplace's operator to equation Eq.27:

$$\Delta\left(\frac{\partial\vartheta_y}{\partial x} - \frac{\partial\vartheta_x}{\partial y}\right) + \Delta\Delta w = -\frac{3}{2Gh}\Delta P \tag{52}$$

or

$$\frac{\partial}{\partial x}(\Delta\vartheta_y) - \frac{\partial}{\partial y}(\Delta\vartheta_x) + \Delta\Delta w = -\frac{3}{2Gh}\Delta P. \tag{53}$$

However, Eqs.51 and 52 yield

$$\frac{\partial}{\partial x}(\Delta\vartheta_y) - \frac{\partial}{\partial y}(\Delta\vartheta_x) = \frac{1}{4}\frac{\partial^2}{\partial x^2}(\Delta w) + \frac{1}{4}\frac{\partial^2}{\partial y^2}(\Delta w) +$$
$$+ \frac{5(1-\mu)}{h^2}\left(\frac{\partial\vartheta_y}{\partial x} + \frac{\partial^2 w}{\partial x^2} - \frac{\partial\vartheta_x}{\partial y} + \frac{\partial^2 w}{\partial y^2}\right) - \frac{\mu h^2}{8D(1-\mu)}\Delta P. \tag{54}$$

According to Eq.27 the last equation can be written

$$\frac{\partial}{\partial x}(\Delta\vartheta_y) - \frac{\partial}{\partial y}(\Delta\vartheta_x) = \frac{1}{4}\Delta\Delta w - \frac{5(1-\mu)}{h^2}\frac{3}{2Gh}P -$$
$$- \frac{\mu h^2}{8D(1-\mu)}\Delta P, \tag{55}$$

and substituting in Eq.°53 we obtain

$$\frac{1}{4}\Delta\Delta w - \frac{5(1-\mu)}{h^2}\frac{3}{2Gh}P - \frac{\mu h^2}{8D(1-\mu)}\Delta P + \Delta\Delta w = -\frac{3}{2Gh}\Delta P. \tag{56}$$

Operating and ordering it gives

$$\Delta\Delta w = \frac{P}{D} + \frac{6(1+\mu)(\mu-2)}{5Eh}\Delta P. \tag{57}$$

The system of governing equations turns out to be

$$\Delta\Delta w = \frac{P}{D} + \frac{6(1+\mu)(\mu-2)}{5Eh}\Delta P \tag{58}$$

$$\Delta\vartheta_x + \frac{5(1-\mu)}{h^2}\vartheta_x = -\frac{1}{4}\frac{\partial}{\partial y}(\Delta\widehat{w}) + \frac{5(1-\mu)}{h^2}\frac{\partial\widehat{w}}{\partial y} + \frac{\mu h^2}{8D(1-\mu)}\frac{\partial P}{\partial y} \tag{59}$$

$$\Delta\vartheta_y + \frac{5(1-\mu)}{h^2}\vartheta_y = -\frac{1}{4}\frac{\partial}{\partial x}(\Delta\widehat{w}) + \frac{5(1-\mu)}{h^2}\frac{\partial\widehat{w}}{\partial x} - \frac{\mu h^2}{8D(1-\mu)}\frac{\partial P}{\partial x} \tag{60}$$

### IV. GOVERNING EQUATIONS DISCONNECTED IN DISPLACEMENTS AND "MOMENTS-SUM"

Now, we propose to transform the system formed by Eqs 5 and 10 by other one disconnected in which displacements and "moments-sum", also called Marcus moment, take part.

We defined "moment-sum" (designated by $M$) to,

$$\frac{M_x + M_y}{(1+\mu)} = M\left(\frac{\partial\hat{\tau}_{xz}}{\partial x} + \frac{\partial\hat{\tau}_{yz}}{\partial y}\right) \tag{61}$$

Adding equations 9 we obtain,

$$\frac{\partial\vartheta_y}{\partial x} - \frac{\partial\vartheta_x}{\partial y} = \frac{M}{D} + \frac{1}{5G}\left(\frac{\partial\hat{\tau}_{xz}}{\partial x} + \frac{\partial\hat{\tau}_{yz}}{\partial y}\right) - \frac{12\mu P}{5Eh} \tag{62}$$

However the sum of the derivatives of the shearing stresses in the middle surface are obtained from the Eqs 6 and 7 of which we deduce

$$\frac{\partial\hat{\tau}_{xz}}{\partial x} + \frac{\partial\hat{\tau}_{yz}}{\partial y} = G\left(\frac{\partial\vartheta_y}{\partial x} + \frac{\partial^2 w}{\partial x^2} - \frac{\partial\vartheta_x}{\partial y} + \frac{\partial^2 w}{\partial y^2}\right) =$$
$$G\left(\frac{\partial\vartheta_y}{\partial x} - \frac{\partial\vartheta_x}{\partial y} + \Delta w\right) \tag{63}$$

And therefore Eq. 62 is transformed into

$$4\left(\frac{\partial\vartheta_y}{\partial x} - \frac{\partial\vartheta_x}{\partial y}\right) = 5\frac{M}{D} + \Delta w - \frac{12\mu P}{5Eh} \tag{64}$$

Also according to Eq. 26,

$$\frac{\partial\hat{\tau}_{xz}}{\partial x} + \frac{\partial\hat{\tau}_{yz}}{\partial y} = \frac{-3}{2h}P \tag{65}$$

And thus Eq. 62 gives

$$\frac{\partial\vartheta_y}{\partial x} - \frac{\partial\vartheta_x}{\partial y} = \frac{M}{D} - \frac{3P}{10Gh} - \frac{12\mu P}{5Eh} \tag{66}$$

Now we may eliminate the difference of the derivatives of the rotations with this last equation and Eq. 27, obtaining

$$\Delta w = -\frac{M}{D} - \frac{6P}{5Gh(1+\mu)}. \tag{67}$$

On the other hand, according to Eq. 54 we deduce

$$\frac{5(1-\mu)}{h^2}\left(\frac{\partial\vartheta_y}{\partial x} + \frac{\partial^2 w}{\partial x^2} - \frac{\partial\vartheta_x}{\partial y} + \frac{\partial^2 w}{\partial y^2}\right) = \Delta\left[\frac{\partial}{\partial x}(\vartheta_y) - \frac{\partial}{\partial y}(\vartheta_x)\right] - \frac{1}{4}\Delta\left[\frac{\partial^2}{\partial x^2}(w) + \frac{\partial^2}{\partial y^2}(w)\right] + \frac{\mu h^2}{8D(1-\mu)}\Delta P \tag{68}$$

The first member according to Eq. 27 is:

$$-\frac{5(1-\mu)}{h^2} \cdot \frac{3h^2}{12D(1-\mu)}P = \frac{-5P}{4D} \tag{69}$$

The first addend of the second member according to Eq. 54, 59 and 60 is:

$$\frac{\Delta M}{D} - \frac{3\Delta P}{10Gh} - \frac{12\mu\Delta P}{5Eh} \tag{70}$$

and the second addend of the second member according to Eq. 59 and 60 is:

$$\frac{-P}{4D} - \frac{3(1+\mu)(\mu-2)}{10Eh}\Delta P \tag{71}$$

Substituting in Eq. 68 one obtains

$$\frac{\Delta M}{D} = \frac{-P}{D} + \frac{3\Delta P}{10Gh} + \frac{12\mu\Delta P}{5Eh} + \frac{3(1+\mu)(\mu-2)}{10Eh}\Delta P - \frac{3\mu}{4Gh}\Delta P \quad (72)$$

After operating and ordering

$$\Delta M = -P - \frac{\mu h^2}{10}\frac{\Delta P}{1+\mu} \quad (73)$$

## V. DISCUSSION OF THE EQUATIONS DEDUCED.

If we analyze the system of equations (58)-(60), we can see that if the terms $-\frac{1}{4}\frac{\partial}{\partial y}(\Delta\hat{w})$ and $\frac{\mu h^2}{8D(1-\mu)}\frac{\partial P}{\partial y}$ are neglected in (58) and (59), these equations coincide exactly with the ones obtained in Reissner Bolle theory.

In addition, as we have pointed out, the value of the stress $\sigma_z$ in this work is identical to the one proposed in the theory of plates of Kromm (1953).

The term $-\frac{1}{4}\frac{\partial}{\partial y}(\Delta\hat{w})$ appears as a consequence of the assumption 2) of this work, which accounts for parabolic distribution of the transverse shear strains through the thickness of the plate.

Similar results have been found in Reddy´s work (1999).

Finally, the terms $\frac{\mu h^2}{8D(1-\mu)}\frac{\partial P}{\partial y}$ and $\frac{\mu h^2}{8D(1-\mu)}\frac{\partial P}{\partial x}$ are a consequence of the first assumption, which is not usually taken into account in the theories of plates and shells but is important for variable loads.

Lastly, equation (67) has the same structure than the one proposed by Reismann (1988), relating displacements and moment- sum, which is

$$\Delta w = -\frac{M}{D} - \frac{P}{K^2GH} \quad (74)$$

It only differs from (74) in the term $(1 + \mu)$.

The possibility of expressing the system (58)-(60) in the compact form (67) offers a wide range of possibilities of obtaining analytical solutions for simply supported plates as it will be seen in the following examples.

Otherwise, a Fourier series solution is always possible, taking into account the appropriate boundary conditions.

## VI. ILLUSTRATIVE EXAMPLES

**Example 1**: A simply supported isotropic rectangular plate subjected to uniformly distributed load. The plate is subjected to the transverse load, $q(x,y)$ on surface $z = -h/2$ acting in the upward z-direction as given below

$$q(x,y) = q_0 \sin\frac{\pi x}{a}\sin\frac{\pi y}{b} \quad (75)$$

**A. Navier Solution**

The following is the solution form for $w(x,y)$ satisfying the boundary conditions given for a plate with all the edges simply supported:

$$w(x,y) = c \sin\frac{\pi x}{a}\sin\frac{\pi y}{b} \quad (75)$$

Where coefficient c can be evaluated after substitution of (58)

$$c = \frac{q_0}{D\pi^4\left(\frac{1}{a^2}+\frac{1}{b^2}\right)}\left[\frac{1}{\left(\frac{1}{a^2}+\frac{1}{b^2}\right)} + \frac{\pi^2 t^2(2-\mu)}{10(1-\mu)}\right] \quad (76)$$

Substituting this coefficient in the displacement field (75)

$$w(x,y) = \frac{q_0 a^4}{D\pi^4\left(1+\frac{a^2}{b^2}\right)^2} \cdot \left[1 + \frac{\pi^2 t^2(2-\mu)\left(1+\frac{a^2}{b^2}\right)}{10(1-\mu)a^2}\right]\sin\frac{\pi x}{a}\sin\frac{\pi y}{b} \quad (77)$$

Which can be reduced to the expression obtained by Timoshenko (1959) in case of $t \rightarrow 0$ and coincides exactly with the expression deduced with the higher order plate theory of Vlasov (1958).

This author can be considered the first one to develop a consistent higher order plate theory, establishing a third-order displacement field that statisties the stress-free boundary conditions on the top and bottom planes of a plate, Reddy (1990).

Similar or identical results to the Vlasov´s theory have been found for moments, rotations and shear forces, but the major advantages have been found by calculating horizontal bending stresses.

In Reissner theory, the maximum horizontal bending stress is

$$\sigma_x\left(x,y,-h/2\right) = -1{,}78\sin\frac{\pi x}{a}\sin\frac{\pi y}{b} \quad (78)$$

The exact solution with the theory of elasticity is,

$$\sigma_x\left(x,y,-h/2\right) = -2{,}12\sin\frac{\pi x}{a}\sin\frac{\pi y}{b} \quad (79)$$

In this work, it has been found,

$$\sigma_x\left(x,y,-h/2\right) = -2{,}01\sin\frac{\pi x}{a}\sin\frac{\pi y}{b} \quad (80)$$

Reducing the error to 6%, approximately.

**Example 2**: Free vibration of a simple supported isotropic rectangular plate.

Applying the principle of d'Alembert to establish dynamic equilibrium equations for the study of transverse oscillations of plates, we need only consider in the equilibrium equation (57) the inertia forces rather than static loads P.

The problem to be solved is,

$$\Delta\Delta w = -\frac{\gamma t}{D}\ddot{w} - \frac{6(1+\mu)(\mu-2)}{5E}\Delta\ddot{w} \quad (81)$$

If the deflection is expressed as,

$$w = [C_1 cos(f\tau) + C_2 sen(f\tau)]U(x,y) \quad (82)$$

Where $f$ is the frequency.

Substituting,

$$\Delta\Delta U = \frac{\gamma t}{D}f^2 U + \frac{6(1+\mu)(\mu-2)\gamma}{5E}f^2\Delta U \quad (83)$$

Boundary conditions are satisfied if we take the solution in the form,

$$U_{mn} = sen\frac{m\pi x}{a} sen\frac{n\pi y}{b} \quad (84)$$

Substituting again,

$$\left(\frac{m\pi}{a}\right)^4 + 2\left(\frac{mn\pi^2}{ab}\right)^2 + \left(\frac{n\pi}{b}\right)^4 - \frac{\gamma t}{D}f^2{}_{mn}$$
$$+ \frac{6(1+\mu)(\mu-2)\gamma}{5E}f^2{}_{mn}\left[\left(\frac{m\pi}{a}\right)^2\right.$$
$$\left.+ \left(\frac{n\pi}{b}\right)^2\right] = 0 \quad (85)$$

And solving,

$$f_{mn} = \frac{\pi^2}{b^2}\left[n^2 + \left(m\frac{b}{a}\right)^2\right]\cdot$$
$$\sqrt{\frac{Et^2}{(1+\mu)\gamma}}\frac{1}{\sqrt{12(1-\mu) - 1{,}2(\mu-2)\frac{t^2\pi^2}{b^2}\left[\left(m\frac{b}{a}\right)^2 + n^2\right]}} \quad (86)$$

The first line of this equation corresponds to the classical solution, Leissa (1973).

In order to compare the results for a square plate, Mindlin (1951), they are referred to the nondimensional frequency parameter λ given by,

$$\lambda = fa^2\sqrt{\frac{\rho h}{D}} \quad (87)$$

Results are shown for different ratios thickness/length (0.01, 0.1 and 0.2) in the following Table 1.

| Mode | h/a=0.01 | h/a=0.1 | h/a=0.2 |
|---|---|---|---|
| 1 | 19.734 | 19.067 | 17.450 |
| 4 | 78.848 | 69.794 | 55.158 |
| Sol.1(Mindlin) | 19.734 | 19.067 | 17.450 |
| Sol.4(Mindlin) | 78.850 | 69.795 | 55.160 |

Table1: Parameter λ of a simply supported square plate

Excellent agreement with the theory, for simply supported plates, has been found.

## V. SUMMARY

We have achieved the system formed by Eqs.58, 59 and 60, and Eq. 73 which have the same order of refinement as the one presented by Muhammad *et al* (1990) for thick plates but with the advantage of following the exposition of the classical technical theories; it constitutes a more refined generalization than the presented one by Reisman (1988) for thick plates, and than the other one presented by Timoshenko and Woinowski (1959).

Besides, assuming that the plane of application of the load is the upper plane of the plate, the first predictions of normal stress depending on the thickness values match those deduced by Kromm theory (1953).

Analytical solutions to this system of equations have been presented for simply supported plates in static and dynamic analysis with excellent results.

In future work we will present more complex analytical solutions for other boundary conditions, and numerical results for those cases in which analytical solutions are not possible.